\documentclass[aps,pra,showpacs,twocolumn,floatfix,nofootinbib,superscriptaddress]{revtex4}
\usepackage{graphicx} 
\usepackage{amsmath}
\usepackage{bm}
\usepackage{dcolumn}

\begin{document}

\title{Calculation of Stark-induced absorption on the $6s6p\,^3\!P_1 - 6s^2\,^1\!S_0$ transition in Hg }

\author{  K. Beloy}
 \affiliation{Physics Department, University of
Nevada, Reno, Nevada  89557, USA}

\author{ V. A. Dzuba}
\affiliation{Physics Department, University of
Nevada, Reno, Nevada  89557, USA}

\affiliation {
School of Physics, University of New South Wales, Sydney,
2052, Australia}

\author{A. Derevianko }
\affiliation{Physics Department, University of
Nevada, Reno, Nevada  89557, USA}

\date{\today}
\begin{abstract}
We carry out relativistic many-body calculations of the Stark-induced absorption coefficient
on the 254-nm  $6s6p\,^3\!P_1
(F=1/2) - 6s^2\,^1\!S_0$ line of $^{199}$Hg atom, the effect considered before by Lamoreaux and Fortson using a simple central field estimate
[Phys.\ Rev.\ A {\bf 46}, 7053 (1992)]. The Stark-induced admixing of states of opposite parity opens additional M1 and E2 transition channels. We find that the resulting
M1-E1 absorption
dominates over E2-E1 absorption. The value of the E2-E1 absorption coefficient depends strongly on the details of treatment of the correlation problem. As a result, our numerical values differ substantially from those of the earlier central field calculation.  Reliable calculation of this effect can enable a useful experimental check on the optical technique being used to search for a permanent electric dipole moment of the $^{199}$Hg atom.
\end{abstract}

\maketitle


\section{Introduction}
The $F=1/2 \rightarrow F=1/2$ electromagnetic transition between two atomic states of opposite
parity has necessarily the electric-dipole (E1) character. However, an application of the
external E-field $\mathcal{E}_s$ breaks the spherical and mirror symmetries
of the atomic Hamiltonian and opens all multipolar transition channels.
To the lowest order in $\mathcal{E}_s$, the transitions are determined by the M1
(magnetic-dipole) and E2 (electric-quadrupole) channels. These effects modify the absorption
coefficient of the atomic sample, the corrective M1 and E2 terms scaling linearly with the
electric field~\cite{HodHecFor91,LamFor92}.

\citet{LamFor92} have focused on a specific setup, relevant for the search of the
permanent electric dipole moment (EDM) of Hg atom~\cite{RomGriJac01,GriSwaLof09} (non-vanishing EDM would violate P- and T-reversal symmetries and be
a clear signature of new physics beyond the standard model of elementary particles).
They considered exciting  the
254-nm  $ 6s^2\,^1\!S_0 \rightarrow 6s6p\,^3\!P_1$ transition
of $^{199}$Hg atom. This isotope has the nuclear spin $I=1/2$. A laser resolves the hyperfine structure
of the $^3\!P_1$ level. It resonantly drives  transitions from a given magnetic $M_{F_i}$ sublevel of the $F_i=1/2$ ground state
 to the $F_f=1/2$ level of the excited state. Then,
for the  $F_i=1/2 \rightarrow F_f=1/2$ transitions, the  relative change in the absorption
coefficient $\alpha$ may be parameterized as~\cite{LamFor92}
\begin{equation}
\frac{\delta\alpha}{\alpha}=\left(  a_{M1}+a_{E2}\right)  ~\left(
\mathbf{\hat{\varepsilon}}_{L}\cdot\mathcal{E}_{s}\right)  \left(
\mathbf{\hat{k}}_{L}~\times~\mathbf{\hat{\varepsilon}}_{L}\right)
\cdot\left(  \frac{\langle\mathbf{F}_{i}\rangle}{F_i}\right)
\label{Eq:relChangeAlpha} \, .
\end{equation}
Here $\hat{\varepsilon}_{L}$ is the polarization vector and $\mathbf{\hat{k}}_{L}$ is
 the direction of propagation of the laser wave. $\langle\mathbf{F}_{i}\rangle$ is  the expectation value of the total angular momentum, i.e., the nuclear polarization in the ground $J_i=0$  atomic state.

In the current $^{199}$Hg EDM experiment~\cite{GriSwaLof09}, the 254-nm transition is used to monitor the nuclear spin direction and thereby detect EDM-induced shifts in nuclear spin precession, which will be linear in an external electric field.  The Stark interference effect on the 254 nm absorption given in Eq.~(\ref{Eq:relChangeAlpha}) also is linear in electric field $\mathcal{E}_{s}$ and depends upon the nuclear spin direction.  A reliable calculation of this effect can enable a useful check on the EDM method when the effect is measured under the same experimental conditions as in the EDM search~\cite{FortsonPrivate}.

\section{Expressions for absorption coefficients}
The goal of this
paper is to compute the atomic-structure coefficients $a_{M1}$ and
$a_{E2}$. One may qualitatively understand the appearance of M1 and
E2 admixtures in Eq.~(\ref{Eq:relChangeAlpha}) as follows. The
Stark-induced transition amplitude in a laser field is composed from
terms linear in the interactions with the external E-field,
$-\mathbf{D}\cdot\mathcal{E}_s$ ($\mathbf{D}$ being the dipole
operator), and the driving $2^k$-pole laser field. We may recouple
the products of the two tensors ($\mathbf{D}$ and the laser EM
multipolar interaction); the resulting compound operators have the
multipolarities of $|k-1|, k, k+1$. For the $F_i=1/2 \rightarrow
F_f=1/2$ transition, $k$ would be limited to 1 and 2. The additional
constraint imposed by the parity selection rule yields the M1 and E2
multipolar couplings.

We derived the expressions for $a_{M1}$ and
$a_{E2}$ using the multipolar expansion of the plane EM wave and the
first-order perturbation theory in the Stark field for the wavefunction.
We employ a
geometry where the quantization axis $\hat{z}$ is chosen along the
$k$-vector of the linearly-polarized laser. The DC Stark-field and
the laser polarization are aligned along the $x$-axis, and the atom
has a definite value of $\hat{F}_y$ in the initial state. This particular
choice of geometry is convenient for working with the most general
relativistic expressions for the multipolar transition
operators~\cite{Joh07book}. By evaluating
Eq.~(\ref{Eq:relChangeAlpha}) of Ref.~\cite{LamFor92} in this
geometry, we identify the following expressions for the structure
factors
\begin{eqnarray}
a_{M1}&=& \sqrt{\frac{2}{3}} \, \frac{R^{M1}}
{\langle n_{i}J_{i}||\mathbf{r} ||n_{f}J_{f}\rangle} \, , \label{Eq:aM1}\\
a_{E2}&=& -\frac{1}{4}\sqrt{\frac{3}{5}} k_L  \, \frac{R^{E2}}{\langle n_{i}%
J_{i}||\mathbf{r}||n_{f}J_{f}\rangle}  \, . \label{Eq:aE2}
\end{eqnarray}
Unless specified otherwise, here and below we use the atomic units, $\hbar=|e|=m_e=1$, and the Gaussian units for EM equations.
Here  ${\langle n_{i}J_{i}||\mathbf{r} ||n_{f}J_{f}\rangle}$ is (within a factor of $-1$ corresponding to the charge of the electron) the conventional reduced dipole matrix
element for the $ 6s^2\,^1\!S_0 \rightarrow 6s6p\,^3\!P_1$
transition and $k_L = \omega_L/c$ is the magnitude of the wavevector
of the laser.
 The quantities $R^{M1}$ and $R^{E2}$ are sums over a
complete set of intermediate states; these sums arise due to the
Stark-induced perturbation and involve the static E1 operator and
multipolar AC couplings to the driving laser field. Each of the
sums, $R^{M1}$ and $R^{E2}$, may be further split into two sums
$S^{M1/E2}_{i/f}$, subscript $i$ or $f$ indicating which of the
states, initial or final, is Stark-perturbed,
\begin{eqnarray}
R^{M1} &=& -S_{i}^{M1}\left(  1^{o}\right)  +S_{f}^{M1}\left(  1^{e}\right) \, ,\\
R^{E2} &=& -\frac{2}{3}\sqrt{2}S_{i}^{E2}\left(  1^{o}\right)
-2\sqrt {\frac{2}{15}}S_{f}^{E2}\left(  2^{e}\right) \, .
\end{eqnarray}
The argument of the sums $S^{M1/E2}_{i/f}(J^\pi)$ indicates the
total angular momentum $J$ and the parity $\pi$ of the intermediate
states as fixed by selection rules. Explicitly, the reduced sums are
\begin{eqnarray}
S_{i}^{M1}\left( J^\pi_n= 1^{o}\right) &=& \sum_{n}\frac{\langle n_{i}J_{i}||\mathbf{r}||n_{n}J_{n}\rangle\langle n_{n}J_{n}||Q^{\left(M1\right)  }||n_{f}J_{f}\rangle}{E_{i}-E_{n}} \, ,\nonumber \\
S_{f}^{M1}\left( J^\pi_n= 1^{e}\right)  &=& \sum_{n}\frac{\langle n_{i}J_{i}||Q^{\left(  M1\right)  }||n_{n}J_{n}\rangle\langle n_{n}J_{n}||\mathbf{r}||n_{f}J_{f}\rangle }{E_{f}-E_{n}}\, , \nonumber \\
S_{i}^{E2}\left( J^\pi_n= 1^{o}\right)   &  = &
\sum_{n}\frac{\langle
n_{i}J_{i}||\mathbf{r}||n_{n}J_{n}\rangle\langle
n_{n}J_{n}||Q^{\left(E2\right)  }||n_{f}J_{f}\rangle}{E_{i}-E_{n}}
\, , \nonumber
\\
S_{f}^{E2}\left( J^\pi_n=2^{e}\right)   &  = & \sum_{n}\frac{\langle
n_{i}J_{i}||Q^{\left(  E2\right)  }||n_{n}J_{n}\rangle\langle
n_{n}J_{n}||\mathbf{r}||n_{f}J_{f}\rangle}{E_{f}-E_{n}}\, .
\nonumber
\end{eqnarray}
We employ the relativistic formalism for the multipolar transition operators $Q^{\left(  M1/E2\right)}$.
Specific single-particle reduced matrix elements computed using Dirac orbital parameterization of Ref.~\cite{Joh07book} are
\begin{eqnarray}
\langle i||Q^{\left( EJ\right)  }||j\rangle &=&  \langle\kappa_{i}%
||C_{J}||\kappa_{j}\rangle
\nonumber \\ \, &
\times&%
\int_{0}^{\infty}r^{J}\{G_{i}(r) G_{j} (r) +F_{i}(r) F_{j}(r)\}dr ,~~~~\\
\langle i||Q^{\left( MJ \right)  }||j\rangle  &=& \frac{\kappa_{i}%
+\kappa_{j}}{J+1}\,\langle-\kappa_{i}||C_{J}||\kappa_{j}\rangle
\nonumber \\
\,
&\times&\int_{0}^{\infty}r^{J}\{G_{i}(r)F_{j}(r)+F_{i}(r)G_{j}(r)\}dr.~~~~
\end{eqnarray}
In both expressions we used the long-wavelength approximation, as $ \alpha \omega_L \ll 1$. In these expressions, $G(r)$ ($F(r)$)
are the large (small) radial components of the Dirac bi-spinor, $\kappa$ are the relativistic angular quantum numbers, and
$C_J(\hat r)$ are the normalized spherical harmonics.

\section{Atomic-structure formalism}
 Mercury atom has two valence
electrons outside a closed-shell core and we start our calculations
with the so-called frozen core ($V^{N-2}$) Dirac-Hartree-Fock (DHF)
approximation. In this approximation, the core orbitals are obtained
self-consistently, while excited (valence) orbitals are subsequently
generated by solving the Dirac equation in the resulting potential
of the core. Such orbitals correspond to the Hg$^+$ valence
orbitals. They are used as a basis for the standard configuration
interaction (CI) technique for two valence electrons (see, e.g.~\cite{LinMor86}). We refer to this approximation as CI-DHF.
Further significant improvement of the accuracy of the calculations
is achieved when the standard CI technique is combined with
many-body perturbation theory (MBPT) to include correlations of the
valence electrons with the atomic core (CI+MBPT).

The CI+MBPT formalism has been discussed in a number of papers (see,
e.g.,~\cite{DzuFlaKoz96,DzuJoh98,BelDerJoh08}). The effective
operator (self-energy, $\hat\Sigma$) arising from the core
polarization may be split into a single-particle, $\hat\Sigma_1$,
and a two-particle, $\hat\Sigma_2$, part acting in the model space.
Qualitatively, a field of the valence electron induces an electric
dipole of the polarizable core:  $\hat\Sigma_1$ describes an
interaction of the valence electron with the self-induced core
dipole, while $\hat\Sigma_2$ describes its interaction with the core
dipole induced by the {\em other} valence electron. We compute the
self-energy correction in the second order of MBPT for the residual Coulomb interaction.
Effects of higher orders will be also included in a semi-empirical fashion, discussed below.


We use the Brillouin-Wigner flavor of MBPT~\cite{LinMor86} to avoid
the ``intruder-state problem'', when the virtual core excitations
inside $\hat\Sigma_2$ become resonant with the states of the valence
subspace. Finally, we emphasize that our computations are {\em ab
  initio} relativistic and employ the Dirac equation and bi-spinors
throughout the entire calculation.

We use the second-order MBPT to calculate the self-energy operators
$\hat \Sigma_1$ and $\hat \Sigma_2$ via direct summation over a
complete set of single-electron states. This set of basis states is
constructed using the B-spline technique~\cite{JohSap86}. We use 40
B-splines of order 9 in a cavity of 40 Bohr radius. The same basis
of the single-electron states is also used in constructing the
two-electron basis states for the CI calculations. We employ partial
waves $\ell=0-4$ and the 14 lowest states above the core in each
partial wave ($s_{1/2}$, $p_{1/2}$, $p_{3/2}$, etc.) for the valence
CI subspace and $\ell=0-5$ and 30 lowest states in each partial wave
for internal summations inside the self-energy operator.

Higher-order correlations are also included in $\hat\Sigma$ in a way
 similar to  Ref.~\cite{DzuFla07}. The
$\hat\Sigma_1$ operator depends on the symmetry of the valence
orbital. Therefore, we have a set of different $\hat\Sigma_1$
operators for $s_{1/2}$, $p_{1/2}$, $p_{3/2}$, etc.\ states. An analysis
of the spectra of Hg (see Table~\ref{Tab:Energies}) shows that
accurate treatment of $\hat\Sigma_1$ is most important for
$s$-electrons, because the ground $6s^2$ state and other states with
$s$ electrons come close to the core and therefore core-valence
correlations must be sizeable for them. In contrast, the core-valence
correlations are much smaller for more diffuse $p$ and $d$ orbitals.
It turns out that the best accuracy is achieved if the all-order
$\hat\Sigma_1^{\infty}$~\cite{DzuFlaSus89} operator is used for the $s$
electrons, the second-order $\hat\Sigma_1$ is employed for the $p$ electrons and
no $\hat\Sigma_1$ is included for $d$ and higher waves.

Higher-order contributions to $\hat\Sigma_2$ are included semi-empirically via
screening factors which modify Coulomb integrals of the second-order
$\hat\Sigma_2$ (see Ref.~\cite{DzuFla07} for details). The values of
these factors are $f_0=0.9$, $f_1=0.72$, $f_2=0.98$, $f_3=1$,
$f_4=1.02$ and $f_5=1.02$.  These values are found from comparing
second-order and all-order $\hat\Sigma_1$.

Finally, we further rescale the $\hat\Sigma_1$ operator for the $s$
and $p$ electrons to fit the experimental spectrum better. The
rescaling coefficients are $\lambda_s=1.0961$ and $\lambda_p=0.8675$.
We use the same $\lambda_p$ for $p_{1/2}$ and $p_{3/2}$ waves.
Note that $\lambda_s>1$ because high-order effects, included in
$\hat\Sigma_1$ for the $s$ electrons, significantly reduce its value. On
the other hand, $\lambda_p<1$ because the second-order MBPT always
overestimates the correlation correction.

The resulting energies  are listed  and compared with experiment in
Table~\ref{Tab:Energies}.
A typical deviation from the experimental values is in the order of 100 $\mathrm{cm}^{-1}$. Even after
the scaling, the disagreement remains, as the number of fitting parameters is limited.

\begin{table}[h]
\caption{Experimental and theoretical energy levels of Hg  (in $\mathrm{cm}^{-1}$).
\label{Tab:Energies}}
\begin{tabular}{llddd}
\hline \hline
\multicolumn{2}{c}{State} & J &
\multicolumn{1}{c}{Exp.~\cite{Moo71}} &
\multicolumn{1}{c}{Theory} \\
\hline
$6s^2$ &  $^1$S     &  0 &     0.000 &   -13.79 \\

$6s6p$ &  $^3$P$^o$ & 0 &  37645.080 & 37458.26  \\
       &            & 1 &  39412.300 & 39312.86  \\
       &            & 2 &  44042.977 & 44265.45  \\

$6s6p$ & $^1$P$^o$  & 1 &  54068.781 & 54180.72  \\

$6s7s$ & $^3$S      & 1 &  62350.456 & 62171.92  \\

$6s7s$ & $^1$S      & 0 &  63928.243 & 63672.24  \\

$6s7p$ & $^3$P$^o$  & 0 &  69516.66  & 69211.87  \\
       &            & 1 &  69661.89  & 69385.18  \\
       &            & 2 &  71207.51  & 70094.95  \\

$6s7p$ & $^1$P$^o$  & 1 &  71295.15  & 71189.34  \\
$6s6d$ & $^1$D      & 2 &  71333.182 & 71295.01  \\

$6s6d$ & $^3$D      & 2 &  71396.220 & 71353.26  \\
\hline \hline
\end{tabular}
\end{table}

The diagonalization of the CI+MBPT Hamiltonian provides us with the
atomic wavefunctions and energies. While the wavefunctions already
have correlation corrections built in, evaluating matrix elements
requires additional inclusion of the so-called screening effect.
This effect arises already in the first order in the residual
Coulomb interaction and describes a re-adjustment of the core
orbitals in response to an externally applied field. We incorporate
the screening in the framework of the all-order many-body technique,
the random-phase approximation (RPA). The RPA formalism (see, e.g., Ref.~\cite{AmuChe75}) describes a linearized
response to an oscillating perturbation. In this regard, while evaluating the reduced sums, we need to fix the driving
RPA frequency for the entire set of matrix elements $Q^{(M1)}$ and
$Q^{(E2)}$ at the photon frequency, $\omega_L=E_{f}-E_{i}$. However,
for the dipole matrix elements (Stark mixing), the RPA frequency
$\omega = 0$.

The evaluation of the sums $S$ requires summing over a complete set
of intermediate atomic states $|n_{n} J_n \rangle$. We use two
approaches: (i) direct summation over states (this implies explicit
computation of the atomic states and evaluation of matrix elements),
and (ii) the Dalgarno-Lewis method. In the Dalgarno-Lewis method~\cite{DalLew55}, the
summation is reduced to solving  the inhomogeneous Schr\"{o}dinger
(Dirac) equation (setup is similar to Ref.~\cite{DerJoh97}). As an
illustration, consider evaluation of the sum $S^{M1}_f$. It may be
represented as
\[
S^{M1}_f = \langle n_i J_i || Q^{(M1)} || \delta \Psi_f \rangle,
\]
where $| \delta \Psi_f \rangle$ lumps corrections to the atomic wave
function of the final state due to the external field. It satisfies
an inhomogeneous equation
\begin{equation}
  (\hat H_\mathrm{eff} - E_f) |\delta \Psi_f \rangle = - \mathbf{r} | n_f J_f \rangle,
\label{eq:H0deltaPsi}
\end{equation}
where $\hat H_\mathrm{eff}$ is the effective CI+MBPT Hamiltonian  of the atom.

\section{ Numerical results}
As an illustration of the CI+MBPT
methodology, we start with calculations of the E1 matrix element and
energy interval for the $6s^2 \,^1\!S_0 -6s6p \, ^3\!P_1$
transition. This matrix element normalizes the Stark-induced
corrections to the absorption coefficient,
Eq.~(\ref{Eq:relChangeAlpha}). The theoretical results at various
levels of approximation and a comparison with the experimental
values are presented in Table~\ref{Tab:3P1-1S0}.
We observe that the core-polarization ($\hat\Sigma_1$) has a
substantial effect on the energy interval, leading to an
improvement in the theory-experiment agreement.
While the
CI+$\hat\Sigma_1$ value of the matrix element perfectly agrees with
the experiment~\cite{CurIrvHen01}, such an agreement is fortuitous:
including the screening correction to the Hamiltonian
($\hat\Sigma_2$) increases its value by a factor of 1.6; only the
additional inclusion of the RPA screening and semi-empirical scaling
moves the theoretical value into a 10\% agreement with  a
2\%-accurate experiment.  We find such an accuracy acceptable, as
{\em ab initio} matrix elements of the intercombination
(spin-forbidden) transitions are known~\cite{PorKozRak01} to be very
sensitive to many-body corrections, the entire values being
accumulated due to the relativistic effects. On the other hand, the
matrix elements of {\em spin-allowed} transitions are stable with
respect to inclusion of the MBPT effects (see, e.g.,
Ref.~\cite{PorKozRak01}). We will return to the evaluation of the
accuracy of our calculations later.

\begin{table}[h]
\caption{Energy interval $\Delta E$ (in $\mathrm{cm}^{-1}$)  and the
reduced electric-dipole matrix element ($R$, a.u.) for the $6s^2
\,^1\!S_0 - 6s6p \, ^3\!P_1$ transition in Hg atom in various
approximations.\label{Tab:3P1-1S0}}
\begin{tabular}{lld}
\hline \hline
Approximation          & $\Delta E$ &
\multicolumn{1}{c}{R} \\
\hline
CI-DHF                   &  31028     &  0.405 \\
CI+$\Sigma_1$            &  37441     &  0.453 \\
CI+$\Sigma_1$+$\Sigma_2$ &  37623     &  0.716\\
CI+$\Sigma_1$+$\Sigma_2$+RPA          &            &  0.577 \\
as above but with all-order $\Sigma_2$     &  36947     &  0.512 \\
as above but with scaled $\Sigma_1$ (Final)     &  39313  &  0.503 \\
\hline
Experiment~\cite{Moo71,CurIrvHen01}
                      &  39412     &  0.453(8) \\
\hline \hline
\end{tabular}
\end{table}

The Stark-induced correction to the absorption coefficient involves two channels, M1 and E2.
We start by discussing the more involved $a_{E2}$ calculations. We
need to compute two sums, $S_{i}^{E2}\left( J^\pi_n= 1^{o}\right)$
and $ S_{f}^{E2}\left( J^\pi_n=2^{e}\right)$. We carry out
calculations (i) by direct summation over the 10 lowest-energy
intermediate states of each symmetry ($1^{o}$ and $2^{e}$) and (ii)
by using the Dalgarno-Lewis method. The latter method is equivalent
to summing over infinitely many intermediate states.
Both calculations use the most
sophisticated CI+MBPT approximation (i.e., 
CI+$\Sigma_1$+$\Sigma_2$+RPA with semi-empirical scaling). The
results are presented in Table~\ref{Tab:SE2}. An examination of
contributions reveals that there are substantial
cancellations inside individual sums. This leads to an enhanced sensitivity
to correlations. For example, consider the value of the $S_{i}^{E2}\left( J^\pi_n= 1^{o}\right)$ sum truncated at the 10 lowest-energy levels.
It changes from
$-35.6$ to $-28.23$ (Table~\ref{Tab:SE2}) while progressing from the CI+$\Sigma_1$+RPA to  the full CI+MBPT  treatment.
 Additional cancellations occur when
the reduced sums are combined into the quantity
$R^{E2}=-\frac{2}{3}\sqrt{2}S_{i}^{E2}\left(  1^{o}\right) -2\sqrt
{\frac{2}{15}}S_{f}^{E2}\left(  2^{e}\right) \approx 24.13 - 32.84 =
-8.71 $. Notice that this value is several times smaller than the
properly rescaled  value of the largest contribution in
Table~\ref{Tab:SE2}. These cancellations may lead to a poor
accuracy of our resulting absorption coefficient
\[
a_{E2} = - 4.39 \times 10^{-3} \ \mathrm{a.u.} = - 0.0853 \times 10^{-8}
/(\mathrm{kV/cm})\, .
\]
This result was obtained using the {\em ab initio} matrix element
from Table~\ref{Tab:3P1-1S0}. Notice that there
is a phase ambiguity  originating from atomic wavefunctions for sums $S$ and the
normalizing dipole matrix element. However, when these quantities are combined in
Eq.~(\ref{Eq:aM1},\ref{Eq:aE2}), the ambiguous phase factors cancel out.
In our particular computation, the sign of the dipole matrix element $\langle 6s^2 \,^1\!S_0 ||\mathbf{r}|| 6s6p \, ^3\!P_1\rangle$  is fixed
by the first entry of Table~\ref{Tab:SE2}.

We proceed to a comparison with results of Ref.~\cite{LamFor92}.
These authors use a simplified approach in which a true
many-electron problem is reduced to a set of single-electron
problems. For the E2 interference they use the Dalgarno-Lewis
summation method based on the  DHF orbitals of the optically active
valence electron. The LS coupling scheme was used in calculations.
Their P-(D-)channel results correspond to our $1^o$ ($2^e$) values.
Unfortunately, Ref.~\cite{LamFor92} contains a number of numerical
mistakes in calculations of $a_{E2}$ coefficient, hindering a
comparison. For example, for the P-channel, using Eq.~(39) of
Ref.~\cite{LamFor92} and their numerical values we obtain, $a_{E2,P}
= -0.96 \times 10^{-8}\, \mathrm{cm/kV} $ which is an order of
magnitude smaller than the published value. Similarly for the
D-channel, based on Eq.~(46) and numerical values of
Ref.~\cite{LamFor92}, we find $a_{E2,D} = 0.1 \times 10^{-8}
\mathrm{cm/kV}$, a factor of 20 smaller than the published value. We
present a detailed comparison with (revised) values of
Ref.~\cite{LamFor92} for the two symmetries of intermediate states
($1^o$ and $2^e$) in Table~\ref{Tab:CompareFortson}. Although of the
same order of magnitude, the individual contributions differ by
signs. The most probable reason for the disagreement is the
sensitivity to particulars of the treatment of correlations. For
example, in computation of the D-channel contribution,
Ref.~\cite{LamFor92} neglected intermediate states of the of $^3D$
symmetry. Hg is a heavy atom, and according to our
table~\ref{Tab:SE2}, omitting the triplet contributions would
increase $a_{E2,D}$ by a factor of three. There is a remarkable
cancellation between individual channels
(Table~\ref{Tab:CompareFortson}): our final result becomes an order
of magnitude smaller than the recomputed value ($a_{E2} = -0.87
\times 10^{-8} \mathrm{cm/kV}$) of Ref.~\cite{LamFor92}.

\begin{table}[h]
\caption{ Breakdown of contributions to the reduced sums for the E2 Stark-induced transition. All
quantities are in atomic units.
\label{Tab:SE2}}
\begin{tabular}{lddd}
\hline \hline
\multicolumn{4}{c}{$S_{i}^{E2}\left( J^\pi_n= 1^{o}\right)$}         \\
\multicolumn{1}{c}{$n_{n} J_n$}& \multicolumn{1}{c}{$\langle 6s^2
\,^1\!S_0||\mathbf{r}||n_{n}J_{n}\rangle$}&
\multicolumn{1}{c}{$\langle
n_{n}J_{n}||Q^{\left(E2\right)}||6s6p\,^3\!P_1\rangle$}&
\multicolumn{1}{c}{contribution}\\
\hline
$6s6p\,^3\!P_1$ & -0.503   &  7.949     & 22.29  \\
$6s6p\,^1\!P_1$ & -2.956   & -4.535     &-54.41  \\
$6s7p\,^3\!P_1$ & -0.037   & -5.460     & -0.63  \\
$6s7p\,^1\!P_1$ &  0.674   & -1.647     &  3.42  \\
$6s8p\,^3\!P_1$ & -0.005   &  1.839     &  0.03  \\
$6s8p\,^1\!P_1$ & -0.286   &  1.652     &  1.35  \\
$6s9p\,^3\!P_1$ & -0.063   & -2.875     & -0.50  \\
$6s9p\,^1\!P_1$ &  0.314   & -0.269     &  0.23  \\
\hline
\multicolumn{3}{l}{Sum(10)}         &   -28.23 \\
\multicolumn{3}{l}{Dalgarno-Lewis, Sum($\infty$)}    &   -25.60 \\
\hline
%
%
%
\multicolumn{4}{c}{$S_{f}^{E2}\left( J^\pi_n= 2^{e}\right)$}         \\
\multicolumn{1}{c}{$n_{n} J_n$}& \multicolumn{1}{c}{$\langle 6s6p
\,^3\!P_1||\mathbf{r}||n_{n}J_{n}\rangle$}&
\multicolumn{1}{c}{$\langle
n_{n}J_{n}||Q^{\left(E2\right)}||6s^2\,^1\!S_0\rangle$}&
\multicolumn{1}{c}{contribution}\\
\hline
$ 6s6d\,^1\!D_2$ &   1.570  &  -6.963  &  75.18  \\
$ 6s6d\,^3\!D_2$ &   2.360  &   3.248  & -52.61  \\
$ 6s7d\,^1\!D_2$ &   0.576  &  -6.153  &  20.67  \\
$ 6s7d\,^3\!D_2$ &   1.354  &   1.661  & -13.10  \\
$ 6s8d\,^1\!D_2$ &   0.263  &  -2.857  &   4.10  \\
$ 6s8d\,^3\!D_2$ &  -1.363  &   0.422  &   3.14  \\
$ 6s9d\,^1\!D_2$ &  -0.284  &   3.654  &   5.47  \\
$ 6s9d\,^3\!D_2$ &  -1.281  &  -0.549  &  -3.71  \\
\hline
\multicolumn{3}{l}{Sum(10)}         &   39.14 \\
\multicolumn{3}{l}{Dalgarno-Lewis, Sum($\infty$)}    &   44.97    \\
\hline \hline
\end{tabular}
\end{table}

\begin{table}[h]
\caption{Comparison of different multipolar contributions to the Stark-induced  absorption coefficients $a_{M1}$ and $a_{E2}$ in
$1/(\mathrm{kV}/\mathrm{cm})$.
The first column gives the character of the multipole, the second column lists values  of Ref.~\cite{LamFor92}, and the third column gives the results of our computation. The notation $a[b]=a \times 10^b$ is used. \label{Tab:CompareFortson} }
\begin{ruledtabular}
\begin{tabular}{ldd}
\multicolumn{1}{c}{Contribution}   &
\multicolumn{1}{c}{Ref.~\protect\cite{LamFor92}}   &
\multicolumn{1}{c}{This work}  \\
\hline
 E2, $1^o$     &  -9.6  [-9]\footnotemark[1]   &      2.4[-9]  \\
 E2, $2^e$     &   1.0   [-9]\footnotemark[1]   &     -3.2[-9]         \\
 E2, total     &  -8.7  [-9]\footnotemark[1]   &      -0.85[-9]  \\[2ex]
 M1, $1^e$      &     0           &    -0.13[-9]            \\
 M1,  $1^o$      &   7.8 [-9]     &   8.9[-9]            \\
 M1, total      &   7.8 [-9]     &   8.86[-9]   \\
\end{tabular}
\end{ruledtabular}
\footnotemark[1]{Values recomputed by us based on data of Ref.~\cite{LamFor92};
 there are errors in numerical evaluations of Eq.(40), Eq.(46), and Eq.(47) of Ref.~\cite{LamFor92}. See text for details. }
\end{table}

Fortunately, while $a_{E2}$ has a poor accuracy, it turns out to be  much smaller
than $a_{M1}$, which, as shown below, can be computed reliably. There are two reduced
sums to evaluate, $S_{i}^{M1}\left(  1^{o}\right)$ and
$S_{f}^{M1}\left(  1^{e}\right)$. Non-relativistically, the
magnetic-dipole operator is diagonal in the radial quantum numbers.
This means that the only substantial contributions arise in the sum
$S_{i}^{M1}\left(  1^{o}\right)$. Indeed, we find from our fully
relativistic analysis
\begin{eqnarray*}
S_{i}^{M1}\left(  1^{o}\right) & \approx & 0.0285 \, , \\
S_{f}^{M1}\left(  1^{e}\right) & \approx & 0.0004 \, .
\end{eqnarray*}
The two dominant matrix elements entering $S_{i}^{M1}\left(
1^{o}\right)$ are $\langle 6s6p\,^3\!P_1 ||
Q^{(M1)}||6s6p\,^3\!P_1\rangle$ and $\langle 6s6p\,^3\!P_1 ||
Q^{(M1)}||6s6p\,^1\!P_1\rangle$. Both matrix elements may be estimated
non-relativistically (e.g., one could use the Land\'{e} formula for the
first matrix element). Further, the term involving the
$|6s6p\,^3\!P_1\rangle$ state is larger by roughly a factor of 5
than the contribution from the singlet state. As a result, the
uncertainty in evaluating  $S_{i}^{M1}\left(  1^{o}\right)$ comes
from the dipole matrix element entering this contribution, the
already discussed $\langle 6s6p\,^3\!P_1 ||
\mathbf{r}||6s^2\,^1\!S_0\rangle$. Incidentally, this is the very
same matrix element that normalizes the absorption coefficient, so
it cancels out in $a_{M1}$. Therefore, with about 25\% accuracy
\[
a_{M1}\approx \sqrt{\frac{2}{3}}\frac{\langle ^{3}P_{1}||Q^{\left( M1\right) }||^{3}P_{1}\rangle }{E_{f}-E_{i}}
\approx 1.19 \times 10^{-8} /(\mathrm{kV/cm})\, ,
\]
where we used the non-relativistic value $\langle
^{3}P_{1}||Q^{\left( M1\right) }||^{3}P_{1}\rangle  = \left(
\frac{3}{4}\right) \sqrt{6}\alpha$. Our full-scale Dalgarno-Lewis
relativistic CI+MBPT calculation results in
\[
a_{M1} = 0.886 \times 10^{-8} /(\mathrm{kV/cm})\, ,
\]
and is consistent  with the non-relativistic estimate. From the
preceding discussion, it is clear that our theoretical value is
stable with respect to neglected many-body corrections.
Ref.~\cite{LamFor92} arrived at the result $a_{M1} = 0.780
\times 10^{-8} /(\mathrm{kV/cm})$. This differs by 12\% from
our estimates.

Finally, we combine the contributions of the M1 and E2
interferences. We note that the poorly known E2 contribution is
fortunately suppressed by a factor of 10 compared to the M1
coefficient.  We find
\[
a_{M1} + a_{E2} = 0.80 \times 10^{-8} /(\mathrm{kV/cm})\, .
\]

\acknowledgements
We gratefully acknowledge discussions with E.N. Fortson and B. Obreshkov.
This work was supported in part by the US National
Science Foundation and by the Australian Research Council.


\end{document}